\documentclass[aps,prb,twocolumn,superscriptaddress,floatfix,longbibliography]{revtex4-1}   
\usepackage{amsmath}    
\usepackage{graphicx}   
\usepackage{hyperref}
\begin{document}
\title{Study of normal modes and symmetry breaking in a
two-dimensional pendulum}
\author{Paramdeep Singh}
\email{chandi@iisermohali.ac.in}
\affiliation{Indian Institute of Science Education \&
Research (IISER) Mohali, Sector 81 Mohali, 140306 India}
\author{R. C. Singh}
\email{rc.phy@gndu.ac.in}
\affiliation{Guru Nanak Dev University Amritsar, Punjab
India 143005}
\author{Mandip Singh}
\email{mandip@iisermohali.ac.in}
\affiliation{Indian Institute of Science Education \& 
Research (IISER) Mohali, Sector 81 Mohali, 140306 India}
\author{Arvind}
\email{arvind@iisermohali.ac.in}
\affiliation{Indian Institute of Science Education \&
Research (IISER) Mohali, Sector 81 Mohali, 140306 India}
\date{\today}
\begin{abstract}
We present an experimental setup to demonstrate normal modes
and symmetry breaking in a two-dimensional pendulum.
In our experiment we have used two modes of a single oscillator
to demonstrate normal modes, as
opposed to two single oscillators used in standard setups of
two-dimensional pendulums.
Breaking of the cylindrical symmetry of the pendulum is
achieved by attaching a spring in the suspension.  This
leads to interesting visual patterns in the motion, wherein
the plane of the oscillator shifts with time, the motion
then becomes elliptical, shifts back again to planar,
before finally
returning to planar motion in 
the original plane.  The symmetry breaking leads to
non-degenerate normal modes of oscillation, whose interplay
gives rise to the observed motion patterns.  This also
explains why for a real pendulum, the plane of motion always
shifts, unlike the ideal two-dimensional pendulum where the
plane of oscillation is supposed to remain fixed.  This
curious fact also contributes to the difficulties involved
in building a Foucault's pendulum, where the plane of
rotation due to Coriolis force needs to be accurately
measured.  The strength of the symmetry breaking in our
system can be quantified by a parameter the ``return time'',
which is defined as the time over which the pendulum returns
to its original motion pattern.  We propose this setup as a
pedagogical tool to introduce the concepts of normal modes
and symmetry breaking in a physics laboratory.  
\end{abstract}
\maketitle
\section{Introduction}
A simple pendulum hanging from a point suspension is in
reality a two-dimensional
pendulum~\cite{peters-teach-99,worland-teach-00,
matthews-jset-01,candela-ajp-01,whitaker-sci-04}. For a long
pendulum length and a small angular displacement, the bob
can be visualized as moving in a two-dimensional horizontal
plane. The oscillation can be started in any direction and the
pendulum will oscillate in the same plane with a frequency
dictated by its length and the  acceleration due
to gravity.  The system has a complete cylindrical symmetry
and therefore oscillations in all planes are completely
identical.

In practice, if one sets up a pendulum, the plane typically
moves in a certain interval of time and the pendulum
acquires an elliptical character to its motion.  These
effects are usually ignored in most experiments as arising
from errors in initial conditions and random perturbations.
In fact, a careful study of such a system reveals that these
effects have a certain systematic character and that the
plane of oscillation of the pendulum moves and after some
time comes back to its original direction in a somewhat
mysterious way~\cite{newburgh-teach-00,greenslade-teach-98}.

It turns out that this is due to the 
non-cylindrically symmetric restoring force arising out of 
imperfections in the suspension. The
symmetry breaking leads to non-degenerate normal modes of
oscillation whose interplay gives rise to these effects.
This is also a major problem while building a Foucault's
pendulum where we would like to have the cylindrical
symmetry to an extent that such effects do not show up even
over several
hours~\cite{criado-ijnm-09,crowell-ajp-81,hart-ajp-87,
jordan-ajp-10,bergmann-ajp-07,rojo-can-09}.

Our aim is to systematically study this phenomenon and
convert it into a pedagogical tool. In order to achieve
this, we explicitly break the symmetry of the pendulum by
attaching a spring with  a very small force constant to the
suspension along one direction. We show how this leads to
the interesting phenomenon of the pendulum becoming
non-planar, being set into elliptical motion, moving into an
entirely different plane and coming back to its original
plane with a certain time period. We demonstrate that this
time period that we call the return time, is associated with
the amount of symmetry breaking in the system and can be
changed by using springs of varying spring constants.

We use a pendulum with a length of  around $2$ m and the
corresponding time period of about $3$ s. The symmetry
breaking springs are chosen such that the symmetry breaking is
small and the associated return times are between $190$ s
and $800$ s. This clear separation of time scales is useful
to ensure that a number of oscillations take place before
any tangible effect of the symmetry breaking sets in.  Since
the effect of the symmetry breaking adds up over the fast
oscillations, it becomes visible eventually.  We also find
it very useful that the pattern of the motion unfolds over
10-15 minutes, allowing a discussion with students around
the experiment. The bob weight is chosen to be about $0.8$
kg so that it stores sufficient energy and allows the system
to perform a large number of oscillations with negligible
damping.  We would like to emphasize that we have used two
coupled modes of a single oscillator in our experiments to
demonstrate the normal modes, which in standard setups is
typically achieved by employing the coupling between two
physical oscillators.

This experiment can be used to teach three concepts: normal
modes, symmetry breaking and appreciating the difficulties
associated with building a Foucault's  pendulum. The
experiment has qualitative as well as quantitative aspects
to it. While different aspects of the experiments can be
demonstrated to a group of students as the the motion
unfolds, there are specific quantitative observations that
the students are expected to make.  They can experimentally
identify normal modes, measure normal mode frequencies and
the return time. The measurements can be  carried out by
hand where one can use a mobile phone as a time measuring
device or one can record the video of the motion and then
use open source tracking software to extract information.

The material in the paper is arranged as follows: In
section~\ref{design_expts} we discuss the design of the
apparatus and the theoretical analysis.
In~\ref{expt_students} we detail the experiments that can be
performed using our setup and we also provide the details of
the measurements taken from the setup.
Section~\ref{discussion} contains conceptual points that can
be brought out through the experiment and can be taken up
with students while they perform the experiment. 
Section~\ref{conc} contain conclusions and future prospects.
\section{Apparatus design and analysis}
\label{design_expts}
The two-dimensional asymmetric oscillator system that we
intend to study is the pendulum oscillating under the effect
of gravity. The frequency can be chosen by adjusting the
length of the pendulum.  The cylindrical symmetry is broken
by attaching a spring toward one side of the suspension so
that if the pendulum is pulled in that direction, apart from
gravity, the spring also provides an additional restoring
force.  The linearity of equations of motion is achieved by
restricting to small angular amplitudes.

\begin{figure}[htbp]
\includegraphics[scale=1]{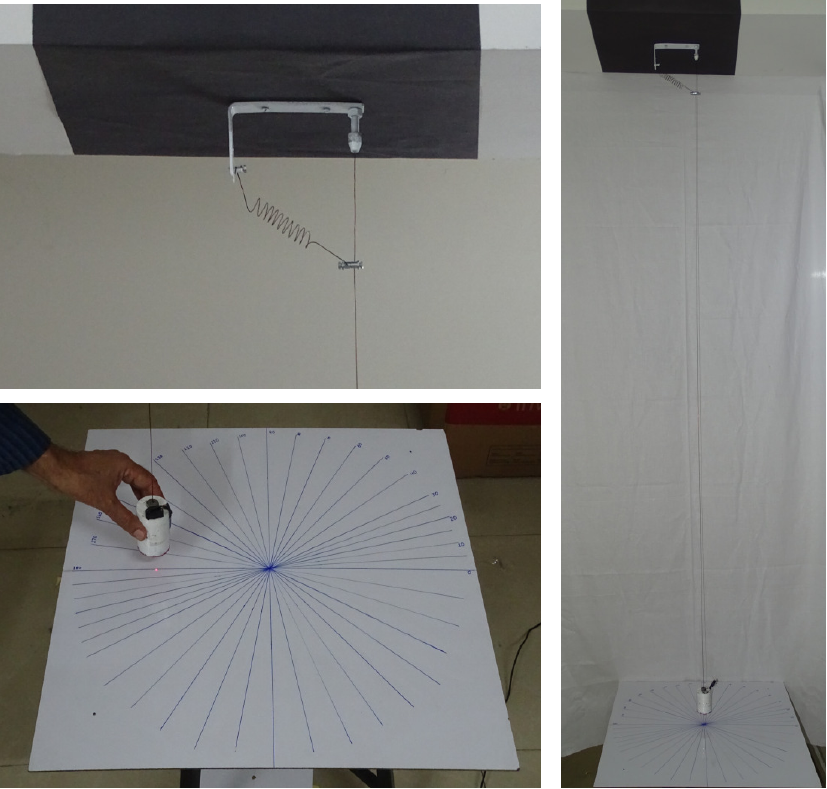}
\caption{\label{pendulum_pics} The top left photograph shows the
suspension of the pendulum with a spring attached to break
symmetry.  The pendulum bob being released at a particular
angle is shown in the photograph on the bottom left. The  
photograph on the right shows the entire setup.}
\end{figure}

A photographs of the actual oscillator setup is depicted in
Figure~\ref{pendulum_pics} in three parts.   The left upper
photograph shows the suspension which is a combination of a
cylindrically symmetric suspension and a spring arrangement
aimed at breaking this symmetry. The left lower photograph
shows the bob and the sheet with angles marked below it. The
right hand side photo is of the entire setup.
  The  length of the system
was  kept fixed at $2$~m. A bob of mass $0.8$~kg was used
and copper wire was used for the suspension. Springs of
different strengths can be attached to change the strength
of symmetry breaking.  A laser pointer was attached below
the bob and the laser spot moves on a sheet of paper
pre-marked with angle markings. We could start the
oscillator by giving it a push in any desired direction.
The system is designed such that the size and the time
periods lead to a visual observation of the motion.  The
time period is about $2.8$~s, which means that we can easily
observe the system and its motion. On the one hand we want
the system to be slow enough so that we can see the motion,
on the other hand the system must perform a large number of
oscillations to achieve a cumulative effect of symmetry
breaking before dissipation becomes dominant.  Spring
constants $k^{\prime}$ of springs have to be chosen such that the
force due to spring is much smaller than gravity. A longer
length also allows one to remain in the linear domain even
for visually large bob amplitudes.  The springs were made by
winding copper wires thicknesses ranging from $0.9$ mm and
$0.56$ mm. on pencils or other cylindrical objects and the
number of turns one winds determine the strength of the
spring. 

\begin{figure}
\label{figure-diag}
\includegraphics[scale=1]{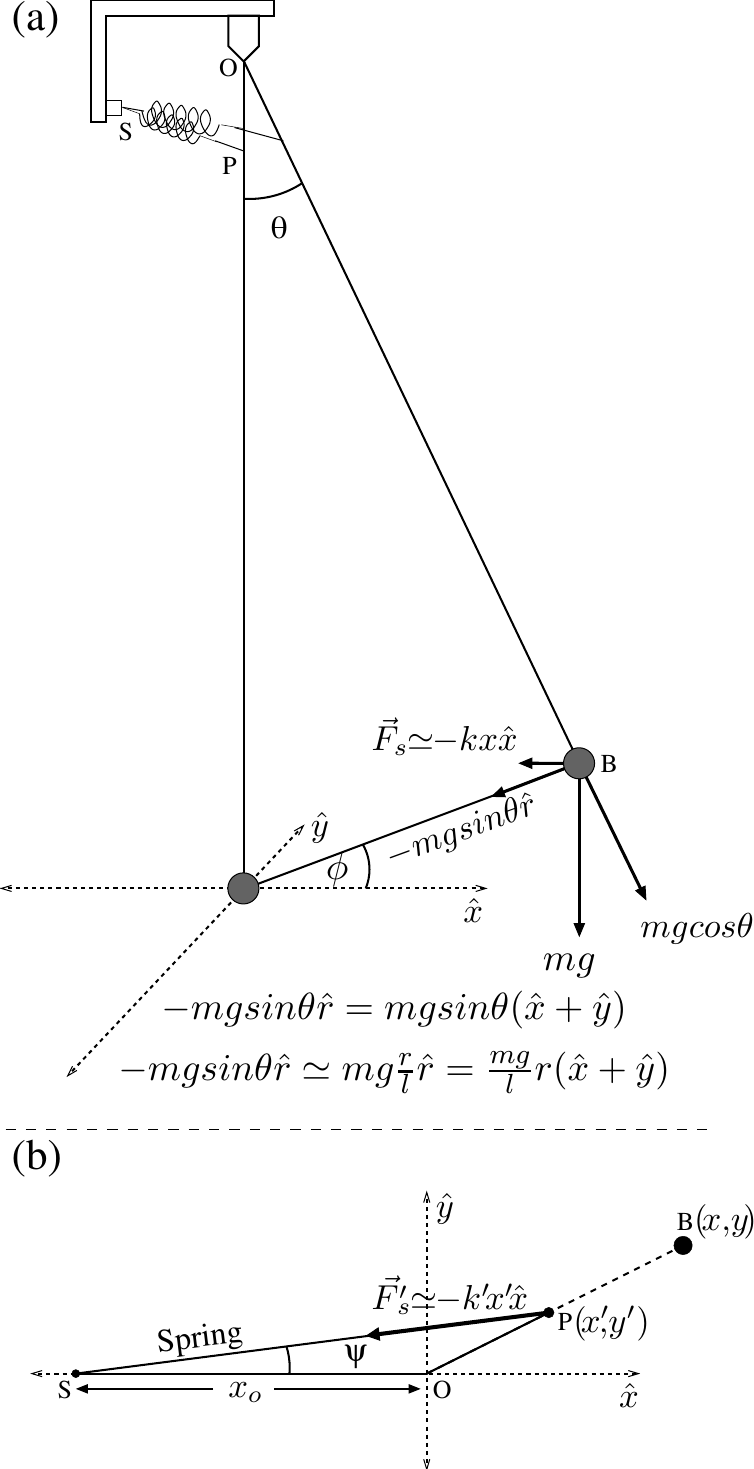}
\caption{(a) Schematic diagram of the pendulum including the
effect of gravitation and spring in the suspension on the
bob.  The bob motion can be considered to happen in the $x-y$
plane for small values and $\theta$ and the effect of
gravity to first order is that of a central linear force in
the $x-y$ plane as shown.  Since the spring is fastened at a
point S in the x-z plane which is a distance $x_0$ away from the point of
suspension O of the pendulum its effect on the bob upto
first order is that of a force proportional to $x$ along
$-\hat{x}$ direction.  (b) The detailed configuration of the
spring  fastened at S and connected to the suspension wire
at P and its force which for small $\psi$
($x_0>>x^{\prime}$) is proportional to $x^{\prime}$ and is
along $-\hat{x}$ direction. The view shown is the projection
in the $x-y$ plane}
\end{figure}

The various forces acting on the pendulum bob are shown
explicitly in the diagram shown in Figure~\ref{figure-diag}.
In the small oscillation approximation when $\theta$ is
small we can consider the pendulum motion in the $x-y$ plain
as shown. In this plane the gravity acts like a central
force and is proportional to the displacement $r$ in the
plane. On the other hand the detail of the force due to
spring is more interesting.  Since the spring is fastened
from a location  in the $x-z$ place at distance 
$x_0$ away from the origin O along the  $-\hat{x}$ direction
and the movement of
the spring is very small compared to the distance $x_0$. One
can easily show that to first order the force due to the
spring is not central in character and is proportional only
to the displacement of the spring along the $\hat{x}$
direction.  This situation is shown in the picture given in
Figure~\ref{figure-diag}(a). The $x-y$ plane projection of 
details of the spring
configuration and its connection to the suspension are shown
Figure~\ref{figure-diag}(b).

The effective Hamiltonian of the bob in terms of its
coordinates $x$ and $y$ and corresponding momenta to first
order can be written as
\begin{eqnarray}
&
H(p_x,p_y,x,y)=\frac{1}{2m} (p_x^2 + p_y^2) + \frac{1}{2} 
(k_1 x^2 + k_2 y^2)& \nonumber \\
&k_1= \frac{mg}{l}+k\quad {\rm and}\quad k_2=\frac{mg}{l}&
\end{eqnarray}
Therefore, the system does not have cylindrical symmetry
and the symmetry is broken because of the presence of the
spring. In case the spring is removed we recover the
cylindrical symmetric situation with $k_1=k_2=k_0$. The amount
of symmetry breaking can be adjusted by using springs of
different spring constants.  This can also be thought of as
if  mass $m$ attached to two springs, one along $x$ and
other along $y$, and each spring is of a different spring
constant.
For the symmetric case when $k_1=k_2=k_0$
If we displace the bob by some distance $d$ in the $x-y$ plane
in any direction and let it oscillate, then the frequency of
oscillation is independent of the direction in which we pull
it, and is given by $w_0=\sqrt{\frac{k_0}{m}}$. The
oscillator moves back and forth in the same direction
forever, without the direction of oscillation changing in
time. Therefore there are an infinite number of linear
polarization modes that we can begin this oscillatory motion
in. We can also start the oscillator in a circular or
elliptical motion and in that case the motion pattern will
again remain the same as time passes, as there is only one
frequency of oscillation in the system. Another way to think
about this system is to think of two independent modes of
the oscillator, one along $x$ and one along $y$, which are
independent and have the same frequency.

What is the motion pattern of such a system
when $k_1\neq k_2$ (the spring is present) and the
symmetry is broken?
Let us write
the equations of motion:
\begin{eqnarray}
m \ddot{x} &=& -k_1 x \nonumber \\
m \ddot{y} &=& -k_2 x 
\label{eoms}
\end{eqnarray}
This means that if we pull the oscillator in the $x$
direction it oscillates with the frequency
$w_1=\sqrt{\frac{k_1}{m}}$ and if we pull it along the  $y$
direction it oscillates with the frequency
$w_2=\sqrt{\frac{k_2}{m}}$. There are two fundamental
frequencies in the system namely $\omega_1$ and $\omega_2$.
Since these motions are decoupled, the motion pattern does
not change and the direction of oscillation remains the same
with the passage of time.

Consider starting the oscillator in a direction which is not
along either $x$ or $y$. What is the motion in this case?
Let us imagine that the oscillator is pulled by an amount
$d$ along  some general direction  
making an angle $\theta$ with respect to the
$x$ axis. This means that the initial condition of the
oscillator  is $X_0=d \cos{\theta}$ and
$Y_0=d\sin{\theta}$.
The equations of motion given in equation~(\ref{eoms})
immediately tell us what the solution is going to be.
The general solution of the equations~(\ref{eoms}) is
\begin{eqnarray}
x(t)&=& A \cos(\omega_1 t + \phi_1)
\nonumber \\
y(t)&=& B \cos(\omega_2 t + \phi_2)
\end{eqnarray}
For the initial conditions that we have started the
oscillator with, the solution reduces to
\begin{eqnarray}
x(t)&=& X_0 \cos(\omega_1 t)
\nonumber \\
y(t)&=& Y_0 \cos(\omega_2 t)
\label{sol_1}
\end{eqnarray}
The displacement in the plane can thus be written as 
\begin{equation}
\vec{r}(t)= X_0 \cos(\omega_1 t) \hat{x} + Y_0 \cos(\omega_2
t) \hat{y}
\label{sol_2}
\end{equation} 
Clearly if $\omega_1=\omega_2=\omega$, the motion will be
oscillatory along a 
straight line and the solution reduces to 
\begin{equation}
\vec{r}(t) = \vec{R}_0\cos(\omega t)
\end{equation}
with $\vec{R}_0 = X_0 \hat{x} + Y_0 \hat{y}$.
However for the case when symmetry is broken
and $\omega_1 \neq \omega_2$ what is the motion pattern?
It turns out that the motion pattern in this case has 
interesting features.  Consider the case when 
the difference in $\omega_1$ and
$\omega_2$ is small. 
Which means that the amount of symmetry breaking is small. In
this case, Equation~\ref{sol_2} implies that  
if we start the oscillator in a straight line
oscillatory motion at an angle $\theta/2$ with the $x$ axis,
after sometime the motion acquires a somewhat elliptical
character with its semi-major axis not along $\theta/2$. The
semi-major axis slowly moves toward $x$ and  the semi-minor
axis slowly grows. After the semi-major axis crosses the $x$
axis, the semi-minor axis begins to decrease and the motion
settles into straight line path again on the opposite side at
an angle $-\theta/2$. From this position the pattern repeats
in the reverse direction and so on.
The motion described above  is in fact a demonstration of
normal modes. There are two normal frequencies in the system
which are $\omega_1$ and $\omega_2$ and the normal mode
motion corresponds to starting the system in either 
the $x$ or the $y$
direction. For such a motion,  the trajectory remains the
same over time. However if we start the oscillator in a
different direction which is not one of the normal modes,
the motion does not remain planar and the motion  pattern
oscillates between two planes on a time scale which depends
upon the coupling strength of the modes represented, via the
difference in frequency of the two normal modes.

When we start the motion in a non normal mode condition,
the time scale over which the direction of oscillation comes
back which we call the $T_R$ to its original direction is 
inversely proportional to
the frequency difference $\Delta \omega=\vert
\omega_1-\omega_2 \vert $ between the modes.  
\begin{equation}
T_R = \frac{2 \pi}{\Delta \omega}
\label{return_time}
\end{equation}
The amount by
which the direction turns accumulates over the oscillation
process. Therefore however small the turning per oscillation
may be, its additive effect over many oscillations is
appreciable and even a small asymmetry will eventually turn
the plane of oscillation of the bob. 
\section{Experiments to be performed}
\label{expt_students}
We propose the following measurements using our
setup, which systematically bring out the various
aspects of the experiment.
\subsection{Motion pattern}
\begin{figure}
\includegraphics[scale=1]{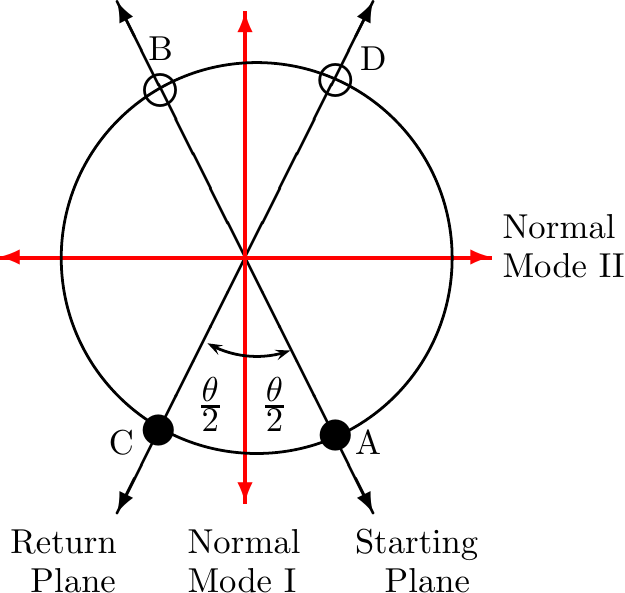}
\caption{The procedure for determination of normal modes.
If the pendulum is started in a planar motion in the plane
AB it oscillates, changes its motion to an elliptical one,
and again settles into a planar motion along CD at a later
time. If we mark these planes on a diagram then we can
compute the normal mode by computing the bisector of the
angle between these planes. The second normal mode is
orthogonal to this direction.}
\label{normal_modes_fig}
\end{figure}
\begin{figure}
\includegraphics[scale=1]{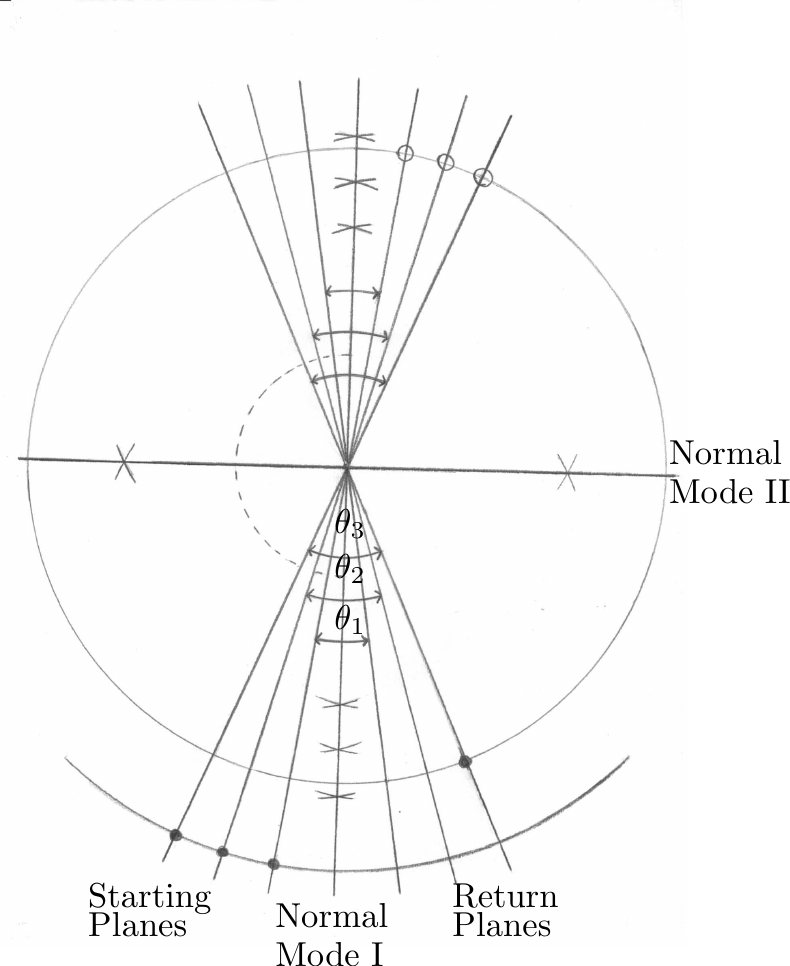}
\caption{Experimental determination of normal modes. The trace
is taken on a plane paper which is scanned and reproduce for
reference. This is a result of the procedure described
in Figure~\ref{normal_modes_fig}}
\label{normal_modes_scan}
\end{figure}
The pendulum is hung with one of the springs attached in the
suspension. The oscillations are started in different
directions the motion is observed alongwith the students.
The qualitative features of the motion are to be documented.
One observes how for a typical initial condition, the bob
starts to oscillate along the direction that it was started
in, the bob then leaves the original direction of motion,
the motion pattern becomes elliptical, then over time the
bob settles into another direction and again motion becomes
elliptical and then returns back to the original direction.
This pattern then repeats itself.  The direction of angular
momentum when the motion is elliptical in the different
phases of the motion is important. It is noticed that
angular momentum flips direction in the second half of the
return cycle, which is indicated by the switch of motion
from clockwise to anticlockwise direction of vice versa. 
This will happen for typical initial
conditions, for two specific directions which are the normal
modes the motion remains planar and does not shift
direction. 
\subsection{Determination of normal modes}
The pendulum is started  in some direction and the direction
is marked on the sheet below. The pendulum then oscillates
and changes its motion pattern as indicated above.  It
finally settles again into linear motion in a different
direction which we again marked on the sheet. This situation
is shown in Figure~\ref{normal_modes_fig}, where we indicate
that if the initial plane of motion is along A-B and the
final plane of motion is along C-D then the pendulum turns
by a total angle of $\theta$ during its motion. The
direction that bisects these two lines is one of the normal
modes of motion. The second normal mode is along the
direction perpendicular to this direction. The normal mode
directions thus computed are indicated in the figure.
Having determined the normal mode directions, we can verify that
indeed we have found the normal modes correctly, by starting
the pendulum along any one  of the normal mode directions, and
observing the motion which then remains planar for a very
long time, without turning at all.  
The actual sheet on which these observations are made is shown
in Figure~\ref{normal_modes_scan} and from this sheet one is
then able to obtain the normal modes. 
\subsection{Return time measurement}
Having determined the normal modes, we know the basic
characteristics of the motion.
As given in Figure~\ref{normal_modes_fig}, if we start the
pendulum along AB, over time the motion becomes elliptical and
finally settles along CD and then returns to the original direction
AB. Return time is defined as the time that the pendulum takes to go
from AB to CD and back to AB.
The return 
time was
measured for different angles and the results are tabulated
in Table~\ref{spring1-2}. 
\begin{table}[htbp]
\begin{center}
\begin{tabular}{|c||c||c|}
\multicolumn{3}{c}{\bf Spring~1}\\
\hline
{Return Angle}  & {Return Time} & {Frequency }\\
{degrees} & {Sec }& {rad/sec}\\
\hline
17    & 192.12 & 0.0327\\
\hline
32    & 192.00 & 0.0327\\
\hline
46    & 192.09 & 0.0327 \\
\hline
\end{tabular}
\end{center}
\caption{\label{spring1-2} Time and frequency data for
the movement of the plane of oscillation of the pendulum.
The time quoted here is the time taken for the  pendulum
started with its plane of oscillation at a certain angle to
return to the same plane after going through the full cycle.
}
\end{table}
Return time turns out to be independent of the angle with
respect to the normal mode from which we release the
bob. The return time in fact is a measure of the amount by
which the symmetry is broken. We will see in the next
subsection how it relates to the normal mode frequencies.
\subsection{Determination of normal mode frequencies}
Next we start the pendulum in each of the normal
modes and measure the time periods of oscillations for
the two normal modes. The results are tabulated in
Table~\ref{table-spring1}.
The time periods and the
associated frequencies will be different. This
difference has developed because of symmetry
breaking and can be related to the extent to which 
symmetry is broken. This difference in frequency
of the two normal modes is directly related to the
return time. 
\begin{table}[htbp]
\begin{center}
\begin{tabular}{|c||c||c||c||c||c|}
\multicolumn{6}{c}{\bf Spring~1}\\
\hline
{ Direction}&{Time for} & {Time for} & {Time for} & {Time
for} & {Time for}\\
{} & {20 Osc} & {40 Osc} & {60 Osc} & {80 Osc} & {100 Osc}\\
\hline
{Mode 1} & 57.21  & 114.38 & 171.54 & 228.78 & 285.98\\
\hline
{Mode 2}& 56.43  & 112.81 & 169.23 & 225.69 & 282.07\\
\hline
\end{tabular}

\vspace*{12pt}
\noindent
\begin{tabular}{|c||c||c|}
\hline
Mode & Time Period (s) & $\omega$ (rad/sec)\\
\hline
Mode 1 & 2.8598$\pm$0.0015 & 2.1971 $\pm$ 0.0011\\
\hline
Mode 2 & 2.8207$\pm$0.0016 & 2.2275 $\pm$  0.0012\\
\hline
\end{tabular}

\vspace*{12pt}
\noindent
\begin{tabular}{|c||c|}
\hline
$\Delta \omega$ & 0.0304 $\pm$0.0023 rad/sec\\
\hline
\end{tabular}
\end{center} 
\caption{\label{table-spring1} 
Measurement of normal mode frequencies after the symmetry is
broken by connecting a spring (No. 1) to  the suspension.}
\end{table}
As mentioned earlier
return time is related to the strength of symmetry breaking
which is reflected in its relationship through the frequency
difference $\Delta \omega$ between the two normal modes. For
this spring the frequency difference suggests an
associated time $\frac{2\pi}{\Delta \omega}=207\pm 15$~s and
the measured return times are close to this number within
experimental error. 
\subsection{Repeating with a set of different springs}
Finally, the entire set of observations is to be repeated for
different springs, i.e.  for different amounts of 
symmetry breaking. With the qualitative observations
remaining the same, the quantitative results change,
indicating a different amount of broken symmetry. We present
the data for two additional springs which are weaker than
the first spring that we used. Table~\ref{table-spring2-2}
contains the return time measurements and 
Table~\ref{table-spring2} contains the normal mode
frequency measurements for the second spring.  
Tables~\ref{table-spring3-2} and~\ref{table-spring3} contain 
similar data for the third spring. 
\begin{table}[htbp]
\begin{center}
\begin{tabular}{|c||c||c|}
\multicolumn{3}{c}{\bf Spring~2}\\
\hline
{Return Angle}  & {Return Time} & {Frequency }\\
{degrees} & {Sec }& {rad/sec}\\
\hline
28    & 459.12 & 0.0137\\
\hline
49    & 462.45 & 0.0136\\
\hline
58    & 464.95 & 0.0136\\
\hline
\end{tabular}
\end{center}
\caption{\label{table-spring2-2} Time and frequency data for
the movement of the plane of oscillation of the pendulum.
The time quoted here is the time taken for the  pendulum
started with its plane of oscillation at a certain angle to
return to the same plane after going through the full cycle.
}
\end{table} 
\begin{table}[htbp]
\begin{center}
\begin{tabular}{|c||c||c||c||c||c|}
\multicolumn{6}{c}{\bf Spring~2}\\
\hline
{ Direction}&{Time for} & {Time for} & {Time for} & {Time
for} & {Time for}\\
{} & {20 Osc} & {40 Osc} & {60 Osc} & {80 Osc} & {100 Osc}\\
\hline
{Mode 1} & 57.24  & 114.43 & 171.68 & 228.94 & 286.15\\
\hline
{Mode 2}& 56.87  & 113.71 & 170.63 & 227.55 & 284.86\\
\hline
\end{tabular}

\vspace*{12pt}
\noindent
\begin{tabular}{|c||c||c|}
\hline
Mode & Time Period (s) & $\omega$ (rad/sec)\\
\hline
Mode 1 & 2.8615$\pm$0.0013 & 2.1958 $\pm$ 0.001\\
\hline
Mode 2 & 2.8446$\pm$0.0017 & 2.2088 $\pm$  0.0013\\
\hline
\end{tabular}

\vspace*{12pt}
\noindent
\begin{tabular}{|c||c|}
\hline
$\Delta \omega$ & 0.013 $\pm$0.0023 rad/sec\\
\hline
\end{tabular}
\end{center} \caption{\label{table-spring2} 
Measurement of normal mode frequencies after the symmetry is
broken by connecting a spring (No. 2) to  the suspension.}
\end{table}
\begin{table}[h]
\begin{center}
\begin{tabular}{|c||c||c|}
\multicolumn{3}{c}{\bf Spring~3}\\
\hline
{Return Angle}  & {Return Time} & {Frequency }\\
{degrees} & {Sec }& {rad/sec}\\
\hline
23    & 849.02 & 0.0074 \\
\hline
32    & 827.66 & 0.0076 \\
\hline
50    & 811.84 & 0.0077\\
\hline
\end{tabular}
\end{center}
\caption{\label{table-spring3-2} Time and frequency data for
the movement of the plane of oscillation of the pendulum.
The time quoted here is the time taken for the  pendulum
started with its plane of oscillation at a certain angle to
return to the same plane after going through the full cycle.
}
\end{table}

\begin{table}[htbp]
\begin{center}
\begin{tabular}{|c||c||c||c||c||c|}
\multicolumn{6}{c}{\bf Spring~3}\\
\hline
{ Direction}&{Time for} & {Time for} & {Time for} & {Time
for} & {Time for}\\
{} & {20 Osc} & {40 Osc} & {60 Osc} & {80 Osc} & {100 Osc}  \\
\hline
Mode 1    & 57.32  & 114.65 & 171.9 & 229.2 & 286.46\\
\hline
Mode 2    & 57.05  & 114.13 & 171.21 & 228.31 & 285.44\\
\hline
\end{tabular}

\vspace*{12pt}
\noindent
\begin{tabular}{|c||c||c|}
\hline
Mode & Time Period (s) & $\omega$ (rad/sec)\\
\hline
Mode 1 & 2.8646$\pm$0.0018 & 2.1934 $\pm$ 0.0014\\
\hline
Mode 2 & 2.8544$\pm$0.0015 & 2.2012 $\pm$  0.0012\\
\hline
\end{tabular}

\vspace*{12pt}
\noindent
\begin{tabular}{|c||c|}
\hline
$\Delta \omega$ & 0.0078$\pm$0.0026 rad/sec\\
\hline
\end{tabular}
\end{center} \caption{\label{table-spring3} 
Measurement of normal mode frequencies after the symmetry is
broken by connecting another spring (No.3) to  the
suspension.}
\end{table}
For the second spring the frequency difference suggests an
associated time $\frac{2\pi}{\Delta \omega}=483\pm 85$~s
which again is close to the measured times and for the third
spring the value $\frac{2\pi}{\Delta \omega}=805\pm 268$~s.
This clearly demonstrates the energy transfer between the
normal modes at a frequency dictated by the frequency
difference between the modes.

It is clear from
the analysis that for a weaker spring when the return time
is long and the difference in the frequency of the normal
modes is small it is difficult to make observations and the
errors are large. It is worth mentioning that our system is
stable and has sufficiently low dissipation so that we are
able to measure the effect of such weak springs. We have
given these results as a demonstration of this fact and the
fact that errors grow when one is trying to measure a small
effect. In actual lab experiment for students one may not
want to use such weak springs.
\subsection{Pendulum base frequency}
As the last measurement, we 
remove the spring and measure the time period
of oscillation of the 
oscillator as it is released from different angles with
respect to the $x$ axis. The data is displayed in 
Table~\ref{no-spring}.
\begin{table}[htbp]
\begin{center}
\begin{tabular}{|c||c||c||c||c||c|}
\hline
{Angle}  & {Time for } &
{Time for} & {Time for}
& {Time for} & {Time
for}\\ {    } & {20
Osc} & {40 Osc} & {
60 Osc} & {80
Osc} & {100 Osc}\\
\hline
$0^{0}$    & 57.29  & 114.52 & 171.83 & 229.04 & 286.28 \\
\hline
$30^{0}$  & 57.18  & 114.38 & 171.69 & 228.94 &  286.19\\
\hline
$60^{0}$  & 57.24  & 114.46 & 171.76 & 229.04 &  286.21\\
\hline
$90^{0}$  & 57.15  & 114.35 & 171.64 & 228.87 &  286.17\\
\hline
$120^{0}$  & 57.37  & 114.58 & 171.83 & 229.11 &  286.35\\
\hline
$150^{0}$  & 57.27  & 114.53 & 171.86 & 229.07 &  286.28\\
\hline
\end{tabular}
\end{center}
\caption{\label{no-spring}
Values of time measured for a cylindrical symmetric
oscillator for different numbers of oscillations and for
different values of the angle of oscillation with respect to
the $x$ axis.}
\end{table}
The frequency of
oscillations can be calculated for different angles 
and are tabulated in Table~\ref{no-spring2}.
\begin{table}[htbp]
\begin{center}
\begin{tabular}{|c||c||c|}
\hline
Angle & Time Period(s) & $\omega$ (rad/s)\\
\hline
0$^{0}$ & 2.8628$\pm$ 0.0021 & 2.1948$\pm$0.0016\\ 
\hline
$30^{0}$ & 2.8619$\pm$ 0.0025 & 2.1955$\pm$0.0019 \\
\hline
$60^{0}$ & 2.8621$\pm$ 0.0029  & 2.1953$\pm$0.0022\\
\hline
$90^{0}$ & 2.8617$\pm$0.0031 & 2.1956$\pm$0.0024\\
\hline
$120^{0}$ & 2.8635$\pm$0.0031 & 2.1942$\pm$0.0023\\
\hline
$150^{0}$ & 2.8628$\pm 0.0025$ & 2.1948$\pm$0.0019\\
\hline
\end{tabular}
\end{center}
\caption{\label{no-spring2} The values of frequency for
oscillation of the pendulum in different directions. The
data shows that the situation is very close to symmetric and
all the polarization modes of oscillation are degenerate
within experimental errors.  Hence the frequency of
oscillation is the same along all directions within
experimental errors. The frequency is $2.1950\pm0.0005$
rad/sec.}
\end{table}
\subsection{Automation aspects}
\begin{figure}[htbp]
\includegraphics[scale=1]{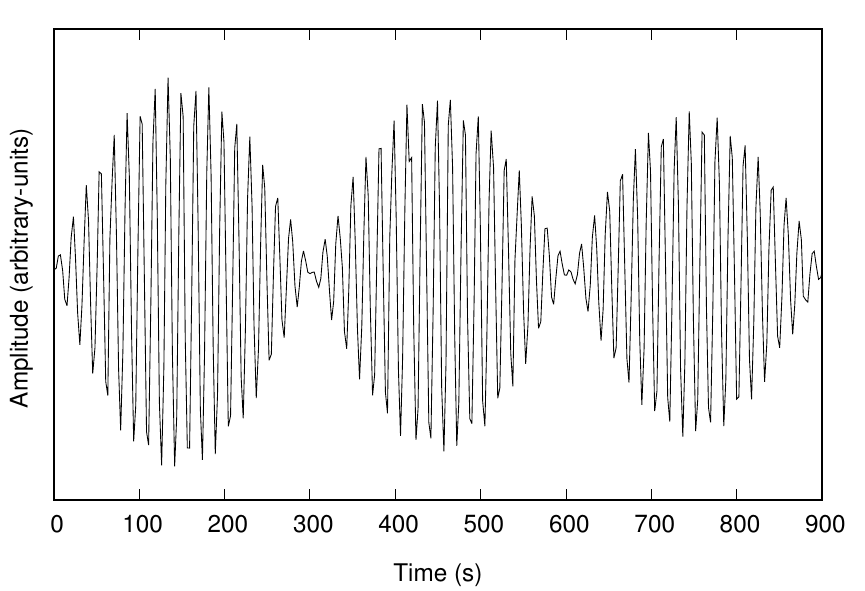}
\caption{\label{automation} Plot of projection of amplitude
along a direction perpendicular to the initial direction
motion as function of time. This data is extracted from the
video and since the data set is large we have plotted only sample
data points. The return time of $300$ s is easily extracted
from the graph. The graph also indicates that the effect of
dissipation is very small even upto $900$ s.}
\end{figure}

The experiments described in the previous section where data
is taken by hand using a stopwatch can also be automated. We
have taken the data in an automated fashion. We used a
webcam to take the video of the motion and then used
tracking software to track the pendulum coordinates as
a function of time during the motion. The data then can be
analyzed at leisure and various aspects of the motion can be
studied. All the four experimental aspects carried out by
hand can be done with this automated setup. 
We analyzed the data on a Linux based system
using the open source physics software
{\em tracker}~\cite{tracker}.
In Figure~\ref{automation} we present a graph
based on such an analysis. In this we have chosen to extract 
the amplitude of motion perpendicular to the initial
direction in which the bob was set in motion. The amplitude 
along the perpendicular direction first increases and then
decreases and finally again comes to zero when the bob
returns to its original oscillation direction. From this
graph we can easily measure the return time. In this case
the return time is $300$ s which reflects the strength of
the spring. The time period along different directions,
normal mode frequencies can also be measured by recording the
corresponding video and extracting data using the tracking
software. The identification of normal modes can also be
done by inspecting the video. 
We have indicated how the entire experiment can be carried
forward and automation aspects can be explored. However we
ourselves found that actually observing and measuring engages
the student better with the experiment than recording a
video and analysis is offline and therefore should not be
used as a substitute for hand measurements. On the other
hand the video provides you with data that can analyzed in
many different ways later. You can ask a question and answer
it by extracting a different aspect of data from the video.
\section{Discussion}
\label{discussion}
In this section we takeup several discussion points that
emerge out of performing the experiment. 
\subsection{Quadratic Hamiltonians}
An interesting point to note here is regarding quadratic
Hamiltonians. A general quadratic Hamiltonian
for a two-dimensional oscillator can be written as
\begin{equation}
H(p_x,p_y,x,y)=\frac{1}{2m} (p_x^2 + p_y^2) + \frac{1}{2} 
(k_x x^2 + k_y y^2+2 k_{xy} xy)
\label{quad-hamiltonian}
\end{equation}

This Hamiltonian has a structure where the kinetic
energy part is invariant under rotation and the potential
energy part can be diagonalized by going to an appropriate
frame which is related to the original frame by a rotation.
Let us look at the potential energy term in quadratic form
\begin{equation}
k_x x^2 + k_y y^2+2 k_{xy} xy
=
\left[\begin{array}{cc} x &  y \end{array} \right]
\left[\begin{array}{cc} k_x &k_{xy}\\k_{xy} &k_y 
\end{array} \right]
\left[\begin{array}{c} x  \\ y \end{array} \right]
\label{quadratic}
\end{equation}
Since the coupling matrix is symmetric, we can always find a
frame in which the matrix is diagonal by performing a
rotation namely 
\begin{equation}
\left[\!\begin{array}{cc} \cos{\theta} &\sin{\theta}
\\-\sin{\theta} & \cos{\theta}
\end{array}\! \right]
\left[\!\begin{array}{cc} k_x &k_{xy}\\k_{xy} &ky 
\end{array} \!\right]
\left[\!\begin{array}{cc} \cos{\theta} &-\sin{\theta}
\\\sin{\theta} & \cos{\theta}
\end{array} \!\right]\! =\!
\left[\!\begin{array}{cc} k^{\prime}_x &0\\0 & k^{\prime}_y
\end{array} \!\right]
\label{rotation}
\end{equation}
The rotation matrix given above corresponds to a rotation of
the coordinate system such that the new coordinates are
given by
\begin{equation}
\left[\begin{array}{c} x^\prime \\ y^\prime 
\end{array} \right]
=\left[\begin{array}{cc} \cos{\theta} &\sin{\theta} \\
-\sin{\theta} & \cos{\theta} 
 \end{array} 
\right]
\left[
\begin{array}{c} x \\ y
\end{array}\right]
\end{equation} 
The novel aspect of our experiment is that this rotation
matrix and the normal modes can be found experimentally! We
start the oscillator in any arbitrary direction, observe its
motion and mark the point of return. The normal mode
direction lies half-way between the starting direction and
final direction from which the oscillator returns. This
allows us to find the angle $\theta$ of the rotation which
diagonalizes the potential energy term and takes us to the
normal coordinates.
\subsection{Attaching several springs}
What if we attach more than one spring in the suspension
pointing in different directions? Will that make the motion
pattern more complicated?  It can be easily done, the best
is to attach several things and observe the motion
pattern experimentally.  It turns out that since spring
forces add like vectors we do not see any difference in the
motion pattern and several springs act like one resultant
spring. Still we have two normal modes and a single return
time. This can be left for the performers to do and figure
out the reasons for it. In other words
the sum total of the
effect of many springs attached in different directions
is represented by a quadratic Hamiltonian of the form
given in equation~(\ref{quad-hamiltonian}). As long as the
springs are linear, there is no further complexity to the
problem, and we can always find the normal modes of the
system  by following the same experimental procedure. In
other words the cylindrical symmetry in two dimensions can
be broken only in one way.
\subsection{Angular momentum conservation}
Another very interesting and peculiar observation one makes
while the pendulum oscillates under broken symmetry (i.e. when
not initially in a normal mode) is that while the pendulum
motion becomes elliptical the sense of the motion also
changes.  The sense of the motion begins as either clockwise
or anti-clockwise (depending upon where the pendulum started)
and after sometime it changes sense during the return path.
This is clearly an indication of the flip of sign of the
angular momentum of the system and implies that the angular
momentum is not conserved. Where does this change in angular
momentum come from and who is applying the torque? A careful
analysis reveals that torque is actually coming from the
suspension and that the angular momentum is in fact
exchanged with the Earth through the suspension!
\subsection{Single two-mode oscillator vs two oscillators}
We have used two modes of a single oscillator and coupled
them in our experiment. In typical setups, where normal
modes are demonstrated and measured, two physical
oscillators are employed. Our experimental design is easier
to implement as  compared to the standard experimental
setups, where if we want to start with two identical
oscillators with the same frequency, it requires tuning,
adjusting the lengths etc.  Another advantage of our setup
is that it also breaks the conceptual barrier in the mind of
a student that we do not need two physical objects for
introducing the coupling and the concept of normal modes, we
basically need two degrees of freedom which can be of the
same physical object.
\subsection{Connection with Foucault's pendulum}
One observes that the symmetry breaking leads to rotation of
the plane of the oscillator over a time scale related to the
strength of symmetry breaking. What if we want to build a
Foucault's pendulum? In that case the oscillator plane
should not shift at least  over one day in the
absence of the Coriolis force. Let us
do an order of magnitude calculation regarding the required
precision for the system. The time period of shift is
related to $\Delta \omega$ and demanding stability over a
period of 24 hours implies that the suspension should be
such that the frequencies should be the same up to parts per
million, in order to observe the effects of Coriolis force!  
\subsection{Connection with Quantum Mechanics}
\label{qm}
The two normal modes of a single pendulum system that we
have presented, where the motion can be a general
superposition of the two modes, is analogous to a two-level
quantum system. Consider a two-level quantum system with the
two eigen states of the Hamiltonian being the stationary
states; they  correspond to the two normal modes of our
pendulum system. If we start the quantum system in one of
the eigen states, the state does not change, which
corresponds to the time invariant motion pattern for the
pendulum when started along either of the normal modes. A
general quantum state will be a linear combination of the
two eigen states and that corresponds to the situation where
we do not start the pendulum in one of the normal modes and
in this case we have a motion pattern which changes with
time. This analogy can be used as a pedagogical tool to
introduce the concept of stationary and non-stationary
states of a quantum system to students. The analogy can be
taken even further if we consider the two-level system to be
a spin-1/2 system with a spin magnetic moment, in this case
it is the magnetic field that breaks symmetry and makes the
energy eigen states non-degenerate. In the absence of a
magnetic field, all the states have same energy,
corresponding to the cylindrical symmetric pendulum and the
switching on of the magnetic field corresponds to the
introduction of the spring in the suspension. 
\subsection{Simulation}
\begin{figure}
\includegraphics[scale=1]{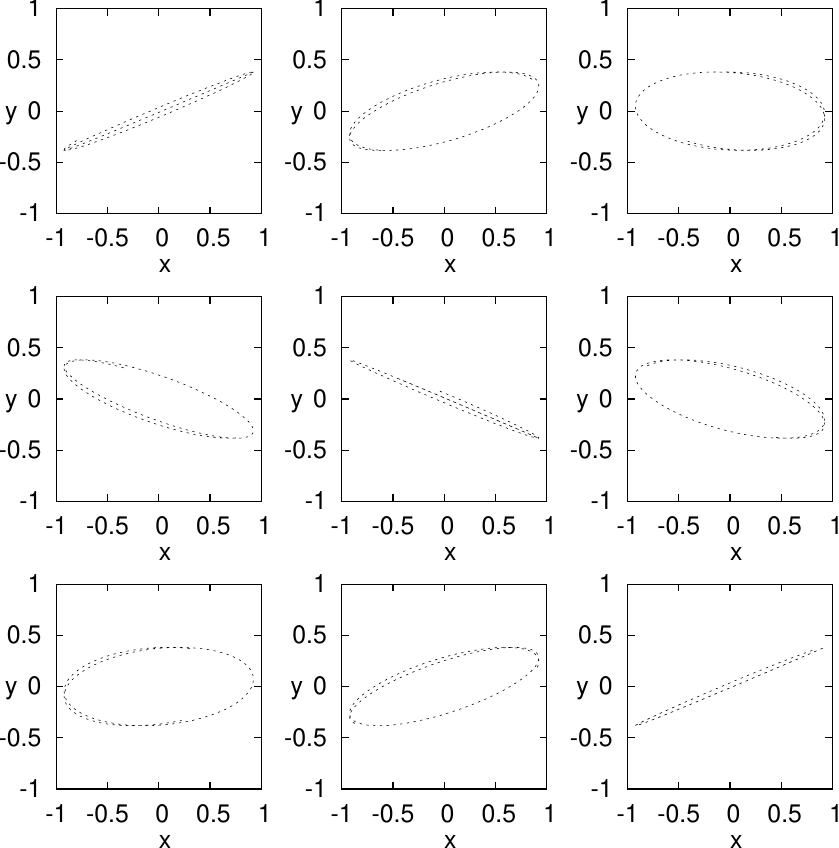}
\caption{\label{pendulum_sim} Simulation of
the motion of a two-dimensional oscillator where $\omega_1
\neq \omega_2$. The difference in frequencies is very
small. The motion is analogous to the experiment.}
\end{figure}

The motion of the pendulum can be simulated where we take
the basic equations given in 
Equations~\ref{sol_1} 
and generate the trajectory on a
computer. In  Figure~\ref{pendulum_sim}
we present the results of such a simulation.
The oscillation is started at an angle of $\pi/8$ with
respect to the $x$ axis. Several snapshots of the motion are
displayed for progressive values of time. In the beginning
the oscillator oscillates in a straight line, with the
passage of time the oscillations  become elliptical and,
the direction of oscillation gradually rotates and the
motion again becomes straight line motion at a certain time
along $-\pi/8$. From this the oscillation slowly comes back
to the original motion pattern.
\section{Concluding Remarks}
\label{conc}
In this paper we have presented an experiment which involves
a two-dimensional pendulum with possibility of breaking the
cylindrical symmetry in a tunable manner. The experiment has
a visual appeal and a group of students can be engaged with
it while demonstrating the different motion patterns that
unfold. The set of experiments that we have described on the
setup involve determination of the normal modes, measuring
the normal mode frequencies and measurement of the return
time. The notion of symmetry breaking in a tunable fashion
is central to the experiment and we demonstrate how even a
small amount of asymmetry in the potential leads to
measurable effects over time as the effects accumulate and
build up. The experiment has many implications that are
taken up in the discussion section and we hope that this
experiment will prove to be an interesting pedagogical tool
for physics laboratories. We have worked with 
several sets of students with our setup in about half a dozen 
institutions and the response has been very good.

In an extension of this experiment, we want to explore this
interplay in detail. We plan to build a suspension which is
non-rigidly connected with the Earth and therefore a part of
it will rotate, compensating the angular momentum gained or
lost by the pendulum. Furthermore, by making the angular
motion sufficiently dissipative, we will kill the angular
part of the motion. In that case the pendulum will be left
with no choice but to settle along one of the normal modes.
This notion can be experimentally connected to the solving
of the eigenvalue problem of a $2 \times 2$ Hamiltonian.
This will mean that if you start the oscillator anywhere, it
will find and settle along an eigenvector of the $2 \times
2$ potential matrix, thereby computing the eigenvector for
you. After the pendulum settles into the normal mode, we can
immediately obtain the corresponding eigenvalue by measuring
its frequency.  The details of this experiment will be
presented elsewhere.
%

\end{document}